\documentclass[a4paper,10pt]{article}
\usepackage[dvips]{graphicx}
\usepackage{color}
\usepackage{amssymb,amsmath}
\numberwithin{equation}{section}

\begin{document}
\title{Einstein-Cartan gravity with scalar-fermion interactions}

\vspace{2cm} 

\author{Olga Razina$^1$,    Yerlan Myrzakulov$^1$,  Nurzhan Serikbayev$^1$,\\ Gulgasyl Nugmanova$^1$,     Ratbay Myrzakulov$^{1,2}$\footnote{The corresponding author. Email: rmyrzakulov@csufresno.edu; rmyrzakulov@gmail.com}\\ \textit{$^1$Eurasian International Center for Theoretical Physics,} \\ \textit{ Eurasian National University, Astana 010008, Kazakhstan}
\\  \textit{$^2$Department of Physics, CSU Fresno, Fresno, CA 93740 USA}}

\date{}

\vspace{2cm}

\maketitle
\begin{abstract} In this paper, we have considered the g-essence and its particular cases, k-essence and f-essence, within the framework of the Einstein-Cartan theory. We have shown that a single fermionic field can give rise to the accelerated expansion within the Einstein-Cartan theory. The exact analytical  solution of the Einstein-Cartan-Dirac equations is found. This solution  describes the accelerated expansion of the Universe with the equation of state parameter $w=-1$ as in the case of $\Lambda$CDM model.
\end{abstract}


\sloppy
\vspace{2cm} 
\section{Introduction}

In the last decades, General Relativity has been extended in  several directions in order to solve some of the problems left open by Einstein's theory in both the ultra-violet and the infra-red regime. One of such extensions is the Einstein-Cartan theory. In particular,  it was proposed that the inflation period  of our  Universe could be described by the spin density which was present in the primordial period of the Universe   within the context of Einstein-Cartan gravity theory  (see e.g. \cite{Ribas}-\cite{Wata2} and references therein). Here we also would like to mention the paper \cite{Guendelman} where was studied exotic low density fermionic states in the two measures field theory. In this work we study the dynamics of  fermion fields within the framework of the Einstein-Cartan theory. For the   f-essence case we construct the simple but not trivial  solutions of the models. 

This work is organized as follows. In section 2  is presented the main properties of the Einstein-Cartan theory.  In the following section 3 we consider the g-essence for the Einstein-Cartan theory case.  The k-essence and the f-essence are presented in sections 3 and 4, respectively. The exact solution of the Einstein-Cartan-Dirac equations is given in section 5. We present our  conclusion in section 6.

\section{Formalism of the Einstein-Cartan gravity}

In this section we briefly review the description of a fermion field coupled to the Einstein-Cartan gravity theory (see e.g.  \cite{Ribas}-\cite{Dereli}). In this case  the affine connection $\Gamma^{\rho}_{\mu\nu}$ has the form
\begin{align}\label{f2}
\Gamma^{\lambda}_{\nu\mu}=\widetilde{\Gamma}^{\lambda}_{\nu\mu}+K^{\lambda}_{\ \nu\mu},
\end{align}
where  $\widetilde{\Gamma}^{\lambda}_{\nu\mu}$ is  the Cristoffel symbol and $K^{\lambda}_{\ \nu\mu}$ is the contortion tensor defined by
\begin{align}\label{f3}
	K^{\lambda}_{\ \nu\mu}=\frac{1}{2}\left(C^{\lambda}_{\ \nu\mu}+C^{\ \ \lambda}_{\nu\mu}+C^{\ \ \lambda}_{\mu\nu}\right).
\end{align}Here the torsion tensor is given by
\begin{align}\label{f1}
C^{\lambda}_{\ \mu\nu}=\Gamma^{\lambda}_{\mu\nu}- \Gamma^{\lambda}_{\nu\mu}\equiv 2\Gamma^{\lambda}_{[\mu\nu]}.
\end{align}
 
  The tetrad $e^a_{\mu}$ is defined as: 
  \begin{equation}g_{\mu\nu}=e^a_{\mu}e^b_{\nu}\eta_{ab},
  \end{equation} where $\eta_{ab}=diag(1,-1,-1,-1)$ is the Minkowski metric tensor.  The tetrad $e^a_{\mu}$   satisfies  the following tetrad condition 
\begin{align}\label{f4}
\mathcal{D}_{\nu}e^{a\mu}\equiv\partial_{\nu}e^{a\mu}+\Gamma^{\mu}_{\rho\nu}e^{a\rho}+\omega_{\nu}^{\ ab}e^{\mu}_b=0,
\end{align}
where $\omega_{\nu}^{\ ab}$ denotes the spin connection. The Riemann tensor is defined as:
\begin{align}\label{f5}
	R^{ab}_{\ \ \mu\nu}=\partial_{\mu}\omega_{\nu}^{\ ab}- \partial_{\nu}\omega_{\mu}^{\ ab}+\omega_{\mu}^{\ ac}\omega_{\nu c}^{\ \ b}- \omega_{\nu}^{\ ac}\omega_{\mu c}^{\ \ b},
\end{align}
and the the Ricci tensor is given by 
\begin{equation}
R_{\mu\nu}=e^{\sigma}_{a}e_{b\nu}R^{ab}_{\ \ \sigma \mu}.
\end{equation}
Here 
\begin{align}\label{f6}
	\omega_{\mu}^{\ ab}=\widetilde{\omega}_{\mu}^{\ ab}+K^{ab}_{\ \ \mu},
\end{align}
where 
\begin{align}\label{f7}
	\widetilde{\omega}_{\mu}^{\ ab}=\frac{e^{a\rho}}{2}(\partial_{\mu}e^b_{\rho}-\partial_{\rho}e^b_{\mu}) -\frac{e^{b\rho}}{2}(\partial_{\mu}e^a_{\rho}-\partial_{\rho}e^a_{\mu}) +\frac{e^{a\rho}}{2}(\partial_{\sigma}e^c_{\rho}-\partial_{\rho}e^c_{\sigma})e^{b\sigma}e_{c\mu}.
\end{align}
    
Note  that 
\begin{align}\label{f8}
	R_{\mu\nu}=\widetilde{R}_{\mu\nu}+\widetilde{\nabla} _{\lambda}K^{\lambda}_{\ \mu\nu}-\widetilde{\nabla}_{\nu}K^{\lambda}_{\ \mu\lambda}+K^{\lambda}_{\ \theta\lambda}K^{\theta}_{\ \mu\nu}- K^{\lambda}_{\ \theta\nu}K^{\theta}_{\ \mu\lambda},
\end{align}
 where the $\widetilde{R}_{\mu\nu}$ and $\widetilde{\nabla}_{\lambda}$ are the Ricci tensor and the covariant derivative referred to the Cristoffel symbol  $\widetilde{\Gamma}^{\lambda}_{\nu\mu}$, respectively. In this paper, we use the following representations for  the Dirac matrices
  \begin{equation}
	\gamma^0=
\begin{pmatrix} I & 0\\
0 & -I\end{pmatrix},\quad \gamma^m=
\begin{pmatrix} 0 & \sigma^m\\
-\sigma^m & 0\end{pmatrix},
\end{equation}
where $I=diag(1,1)$.

\section{G-essence}
Here we consider  the Einstein-Cartan gravity with the g-essence Lagrangian. Its  action reads as
\begin{equation}
	S=\int d^4x e \left[-\frac{1}{16\pi G}e^{\mu}_{a}e^{\nu}_{b}R^{ab}_{\ \ \mu\nu}+K(X,Y,\phi, \psi, \bar{\psi})\right], 
\end{equation}
where $e = det(e^{\mu}_a)$ and  $G$ is the gravitational constant. For simplicity,  we assume that the Lagrangian  of the G-fields has the form
\begin {equation}
K=\alpha_1 X+ \alpha_2 X^n+\alpha_3 V_{1}(\phi)+ Y-mu- V_2(\bar{\psi}, \psi),
\end{equation} 
where $\alpha_j, \beta_j$ are some real constants, $u=\bar{\psi}\psi$ and
\begin {equation}
X=0.5g^{\mu\nu}\nabla_{\mu}\phi\nabla_{\nu}\phi, \quad Y=0.5i(\bar{\psi}\gamma^{\mu}D_{\mu}\psi-\bar{D}_{\mu}\bar{\psi}\gamma^{\mu}\psi)
\end{equation}
are  the canonical kinetic terms, $\nabla_{\mu}$ is the covariant derivative associated with metric $g_{\mu\nu}$. For the Friedmann-Robertson-Walker  metric 
\begin {equation}
ds^2=dt^2-a^2(t)d{\bf x}^2,
\end{equation} 
 the equations of the model have the form
\begin{eqnarray}
	3H^2-8\pi G\rho &=&0,\\ 
		2\dot{H}+3H^2+8\pi Gp&=&0,\\
		\ddot{\phi}+[3H+(\ln{(\alpha_1+\alpha_2n X^{n-1})})_{t}]\dot{\phi}-\frac{\alpha_3V_{1\phi}}{\alpha_1+\alpha_2n X^{n-1}}&=&0,\\
\dot{\psi}+1.5H\psi+im\gamma^0\psi+i\gamma^{0}\frac{dV_2}{d\bar{\psi}}-3\pi Gi\gamma^0(\bar{\psi}\gamma_5\gamma^{i}\psi)(\gamma_5\gamma_{i}\psi)&=&0,\\
\dot{\bar{\psi}}+1.5H\bar{\psi}-im\bar{\psi}\gamma^0 -i\frac{dV_2}{d\psi}\gamma^{0}+3\pi Gi(\bar{\psi}\gamma_5\gamma^{i}\psi)(\bar{\psi}\gamma_5\gamma_{i})\gamma^0&=&0,\\
\dot{\rho}+3H(\rho+p)&=&0.
	\end{eqnarray} 
Here the total energy density $\rho$ and the pressure $p$ of the sources are given by
	\begin{eqnarray}
	\rho &=&\alpha_1X+\alpha_2(2n-1)X^n-\alpha_3V_1+m\bar{\psi}\psi+V_2-\frac{3\pi G}{2}\sigma^2,\\ 
p&=&\alpha_1 X+ \alpha_2 X^n+\alpha_3 V_{1}+\frac{\bar{\psi}}{2} \frac{dV_2}{d\bar{\psi}}+\frac{dV_2}{d\bar{\psi}}\frac{\psi}{2}-V_2-\frac{3\pi G}{2}\sigma^2.
	\end{eqnarray}
	The seach for exact solutions of the coupled system of  differential equations (3.5)-(3.10)  is a very hard job. So we omit this case (see e.g. \cite{MR2}-\cite{MR4} for  General Relativity case).

\section{K-essence }
 If in the action (3.1) we set
 \begin {equation}
K=K_1(X, \phi),
\end{equation}
 then we get the following action for  the k-essence scalar field $\phi$ minimally coupled to the gravitational field $g_{\mu\nu}$   \cite{Mukhanov1}-\cite{Fabbri}
\begin {equation}
S=\int d^{4}x e \left[-\frac{1}{16\pi G}e^{\mu}_{a}e^{\nu}_{b}R^{ab}_{\ \ \mu\nu}+K_1(X, \phi)\right].
\end{equation}

\section{F-essence }

The action of the f-essence we write as  \cite{MR2}
\begin{align}\label{f11}
	S=\int d^4x e \left[-\frac{1}{16\pi G}e^{\mu}_{a}e^{\nu}_{b}R^{ab}_{\ \ \mu\nu}+K_2(Y, \psi, \bar{\psi})\right].
\end{align}

In this section we consider the case \cite{Ribas}-\cite{Wata2}
\begin{align}\label{f9}
K_2=Y-mu-V_2(\bar{\psi},\psi), 
\end{align}
where $\psi$ and $\overline{\psi}=\psi^{\dagger}\gamma^{0}$ denote the spinor field and its adjoint, respectively, $m$ is the fermion mass and $V$ its potential of self-interaction, $u=\bar{\psi}\psi$. The covariant derivatives $D_{\mu}\psi$ and $\overline{D}_{\mu}\bar{\psi}$ are given in terms of the spin connection $\omega_{\mu}^{\  ab}$ by
\begin{align}\label{f10}
	D_{\mu}\psi=\partial_{\mu}\psi+\frac{1}{8}\omega_{\mu}^{\ ab}\left[\gamma_a,\gamma_b\right]\psi, \ \ \bar{D}_{\mu}\overline{\psi}=\partial_{\mu}\overline{\psi}-\frac{1}{8}\omega_{\mu}^{\ ab}\overline{\psi}\left[\gamma_a,\gamma_b\right].
\end{align}

The field equations are obtained from the action \eqref{f11} as follows. The variations of the action (5.1) with respect to the tetrad, $\psi, \bar{\psi}$ and the spin connection imply the following equations, respectively, (see e.g. \cite{Ribas}-\cite{Wata2})
\begin{eqnarray}
	R_{\mu\nu}-\frac{1}{2}Rg_{\mu\nu}-8\pi G T_{\mu\nu} &=&0,\\ 
		i\gamma^{\mu}D_{\mu}\psi-m\psi-\frac{dV_2}{d\bar{\psi}}&=&0,\\
	i\bar{D}_{\mu}\bar{\psi}\gamma^{\mu}+m\bar{\psi}+\frac{dV_2}{d\psi}&=&0,\\
	C^{\mu}_{\ \kappa\lambda}+4\pi G\epsilon_{abcd}e^a_{\lambda}e^b_{\kappa}e^{c\mu}(\bar{\psi}\gamma_5\gamma^d\psi)&=&0,
\end{eqnarray} 
where  $\epsilon_{abcd}$ is the Levi-Civita tensor, $C^{\mu}_{\ \kappa\lambda}$ is the torsion tensor, the energy-momentum tensor is given by
\begin{align}\label{f13}
T_{\mu\nu}=\frac{i}{2}(\bar{\psi}\gamma_{\nu}D_{\mu}\psi-\overline{D}_{\mu}\bar{\psi}\gamma_{\nu}\psi)-\mathcal{L}g_{\mu\nu}. 
\end{align}
Note that the corresponding  contortion tensor is given by
 \begin{align}\label{f16}
		K^{\lambda}_{\ \nu\mu}=-2\pi G\epsilon_{abcd}e^a_{\mu} e^b_{\nu}e^{c\lambda}(\bar{\psi}\gamma_5\gamma^d\psi).
\end{align}
 
We now consider the Einstein-Cartan gravity with the spinor field in  a homogeneous and isotropic Universe described by the spatially flat FRW metric (3.4).
 Then the equations of the model take the form \cite{Ribas}-\cite{Wata2}
\begin{eqnarray}
	3H^2-8\pi G\rho &=&0,\\ 
		2\dot{H}+3H^2+8\pi Gp&=&0,\\
		\dot{\psi}+1.5H\psi+im\gamma^0\psi+i\gamma^{0}\frac{dV_2}{d\bar{\psi}}-3\pi Gi\gamma^0(\bar{\psi}\gamma_5\gamma^{i}\psi)(\gamma_5\gamma_{i}\psi)&=&0,\\
\dot{\bar{\psi}}+1.5H\bar{\psi}-im\bar{\psi}\gamma^0 -i\frac{dV_2}{d\psi}\gamma^{0}+3\pi Gi(\bar{\psi}\gamma_5\gamma^{i}\psi)(\bar{\psi}\gamma_5\gamma_{i})\gamma^0&=&0,\\
\dot{\rho}+3H(\rho+p)&=&0,
	\end{eqnarray} 
where\begin{align}\label{f20}
	\rho=mu+V_2-\frac{3\pi G}{2}\sigma^2, \ \ p=\frac{\bar{\psi}}{2} \frac{dV_2}{d\bar{\psi}}+\frac{dV_2}{d\bar{\psi}}\frac{\psi}{2}-V_2-\frac{3\pi G}{2}\sigma^2, \ \ \sigma^2=(\bar{\psi}\gamma_5\gamma_d\psi)^2.
	\end{align}
	Using the methods e.g. of \cite{Odin}-\cite{MR1}, let us now construct the solution of this system for simplicity assuming 
	\begin{equation}
	\psi=
\begin{pmatrix}	\psi_0 \\ 0 \\ \psi_2 \\ 0 \end{pmatrix}.
\end{equation} Then its simplest but not trivial solution has the form
	\begin{eqnarray}
a&=&	e^{\sqrt{\dfrac{8 \pi G}{3}V_0}\ t+A},\\ 
\psi_0&=&K_0e^{i\kappa-\sqrt{6 \pi GV_0}\ t-A},\\	
\psi_2&=&K_2e^{i\kappa-\sqrt{6 \pi GV_0}\ t-A},
	\end{eqnarray}
where  $A, K_j, V_0, \kappa$ are some real constants. The corresponding potential is given by\begin{equation}
	V_2=V_0-mu-\frac{3\pi G}{2}u^2.
\end{equation}
Note that for this solution the density of energy and the pressure take the form
\begin{equation}
	\rho=V_0, \ \ \ p=-V_0
\end{equation}
so that the equation of state parameter is $w=-1$.

\section{Conclusions}
In this paper, we have considered the g-essence and its particular cases k-essence and f-essence for the Einstein-Cartan gravity theory case. We have shown that a single fermionic field can give rise to the accelerated expansion within the Einstein-Cartan theory. The simplest solution of the Einstein-Cartan-Dirac equations is found. This solution  describes  the accelerated expansion of the Universe with the equation of state parameter $w=-1$.

\end{document}